\documentclass[fleqn,usenatbib]{mnras}

\usepackage{newtxtext,newtxmath}

\usepackage[T1]{fontenc}
\usepackage{ae,aecompl}

\usepackage{graphicx}	
\usepackage{amsmath}	
\usepackage{amssymb}	

\title[Micro-image mags. near critical curves]{Magnifications of paired micro-images emerging from a micro-lensing critical curve}

\author[Weisenbach, Schechter, and Wambsganss]{
Luke Weisenbach,$^{1,2,}$\thanks{E-mail: weisluke@alum.mit.edu}
Paul Schechter,$^{1,3}$
and Joachim Wambsganss$^{2,4}$
\\
$^{1}$Department of Physics, Massachusetts Institute of Technology, 77 Massachusetts Avenue, Cambridge MA 02139\\
$^{2}$Astronomisches Rechen-Institut, Zentrum f{\"u}r Astronomie der Universit{\"a}t Heidelberg, M{\"o}nchhofstr. 12-14,  69120 Heidelberg, Germany\\
$^{3}$MIT Kavli Institute for Astrophysics and Space Research, Cambridge MA 02139\\
$^{4}$International Space Science Institute (ISSI), Hallerstrasse 6, 3012 Bern, Switzerland
}

\date{Accepted XXX. Received YYY; in original form ZZZ}

\pubyear{2019}
\begin{document}
\label{firstpage}
\pagerange{\pageref{firstpage}--\pageref{lastpage}}
\maketitle

\begin{abstract}
Studies of the inner regions of micro-lensed AGN during caustic crossing events have often relied upon the approximation that the magnification near a fold caustic is inversely proportional to the square root of the source-caustic distance. We examine here the behavior of the individual micro-images (one a micro-minimum, the other a micro-saddle) that emerge as a point source crosses a micro-fold caustic. We provide a variety of statistics on both the behavior of the two newly created micro-images, and some parameters which appear in higher order approximations for the magnification. We compare the predictions of these higher order approximations to the actual image magnifications of our simulations.  
\end{abstract}




\section{Introduction}
\label{sec:intro}

The bending of light by gravitational fields can both magnify and create multiple images of an object \citep{1986ApJ...310..568B, 1996astro.ph..6001N}. The gravitational field of a galaxy creates `macro-images' of a distant source such as a quasar. Coherent trends in the magnifications of all the macro-images are due to intrinsic source variability. But, these macro-images are each broken into unresolvable `micro-images' by the individual stars in a galaxy \citep{1979Natur.282..561C, 1986ApJ...301..503P}. Uncorrelated fluctuations among the lightcurves of the macro-images are due to changes in the magnifications of individual micro-images. Of particular interest in the study of gravitationally lensed quasars are caustic crossings, events where the number of micro-images changes by two and the source becomes highly magnified. Such events can, e.g., reveal information about the size and profile of the light emitting region \citep{1991AJ....102..864W, 2018MNRAS.475.1925T}. 

A source crossing a fold caustic is accompanied by the creation or annihilation of a pair of images somewhere along a critical curve. One of these images is a saddlepoint of the light travel time, while the other is a minimum. Together, the newly created images dominate the total magnification (which also contains contributions due to other micro-images) near the caustic. A Taylor expansion of the lens equation in the vicinity of a critical curve allows one to find approximations for the magnifications of these two new, bright, images. 

To leading order, the magnification $\mu$ of a point source near a fold caustic can be approximated as proportional to some flux factor $K$ (also sometimes called the caustic strength), and as inversely proportional to the square root of the source distance normal to the caustic $d$ \citep{1984A&A...130..157C,1989A&A...221....1K,2002ApJ...574..970G}, i.e. \begin{equation}
    \mu=\frac{K}{\sqrt{d}}.
    \label{eq:std-approx}
\end{equation} 

This approximation (that of the `straight fold caustic') is commonly used throughout the literature. Studies of caustic crossing events in the lightcurves of micro-lensed AGN often rely on the convolution of some source luminosity profile with this approximation for the magnification of a point source \citep{1999MNRAS.302...68F,2012MNRAS.423..676A,2015ApJ...814L..26M}. 

To be more precise, it is the sum of the magnifications of the two newly created images which is approximated -- this leading order expression works for each image individually as well, with an appropriate factor of $1/2$. However, while saddlepoint magnifications can take on any value, minima are required to be of unit or higher magnification \citep{1986ApJ...310..568B}. One therefore expects the approximation $\mu=K/\sqrt{d}$ to break down at sufficiently large distances ($d_{break}\approx K^2$).\footnote{If each of the two new images is expected to contribute roughly half of the magnification for a source near a caustic, then there should actually be a breakdown approximately when $\mu_{minimum}=1$, i.e. when $\mu=\mu_{minimum}+\mu_{saddle}=2\rightarrow d=\frac{K^2}{4}$.}

Few authors have considered alternative approximations for the image magnifications. \citet{1999MNRAS.302...68F} derive a similar leading order approximation that takes into account the curvature of the caustic, giving the so-called `parabolic fold caustic' approximation. \citet{2005ApJ...635...35K} and \citet{2011MNRAS.417..541A} have derived higher order approximations for the magnifications. Such higher order approximations necessarily introduce more parameters than $K$ in order to describe caustic crossing events. These other parameters have their own uncertainties as well when measured, making it more difficult to determine the properties of interest (e.g. source size). It is understandable then why many authors work only to leading order, but clearly there are inadequacies in doing so (be it that the approximation is only valid for a small portion of the regime where it is commonly applied, or that it fails to keep the micro-minima above unit magnification). 
    
In what follows we examine the behavior of individual micro-image magnifications near critical curves for a point source near fold caustics, for the parameters of QSO 2237+0305 (Huchra's lens, \citealt{1985AJ.....90..691H}). Our results show clear deviations from the leading order approximation by the time $d=K^2$ is reached. We examine two higher order approximations from \citet{2005ApJ...635...35K} and \citet{2011MNRAS.417..541A} and compare these approximations to the actual image magnifications. We provide some statistics on the behavior of the micro-images (namely, the micro-minima) in our study. Additionally, we present statistics on the parameters present in the higher order approximations with some discussion. Finally, we briefly show how the `shape profile' of a uniform disk crossing a fold caustic is altered under one such higher order approximation.

\section{Simulation setup}
\label{sec:sim_setup}

We consider a micro-lensing star field such that the lens equation in the vicinity of a macro-image relating the source position $\mathbfit{y}$ and image position $\mathbfit{x}$ takes the form \begin{equation}
    \mathbfit{y}=\begin{pmatrix}1-\kappa_{s}+\gamma&0\\0&1-\kappa_{s}-\gamma\end{pmatrix}\mathbfit{x}-\theta_{E}^{2}\sum_{i=1}^{n}m_{i}\frac{(\mathbfit{x}-\mathbfit{x}_i)}{|\mathbfit{x}-\mathbfit{x}_i|^2},
    \label{eq:lenseq}
\end{equation} where the $\mathbfit{x}_i$ are the positions of the $n$ stars with masses $m_i$ (measured in units of some mass $M$ that determines the size of the Einstein ring $\theta_E$).

The shear $\gamma$ is due to the mass distribution of the rest of the galaxy far away from the macro-image. The total surface mass density $\kappa$ is comprised of a smooth component, $\kappa_{s}$, and a portion due to compact matter $\kappa_{\star}$. We take all of our surface mass density $\kappa=\kappa_{s}+\kappa_{\star}$ to be distributed in compact objects, so that $\kappa=\kappa_{\star}$ and $\kappa_{s}=0$. Additionally, we let $\theta_E$ be our unit distance in the image plane, and we let all of our objects be of the unit mass that determines this distance. Equation (\ref{eq:lenseq}) is then more cleanly written as

\begin{equation}
    \mathbfit{y}=\begin{pmatrix}1+\gamma&0\\0&1-\gamma\end{pmatrix}\mathbfit{x}-\sum_{i=1}^n\frac{(\mathbfit{x}-\mathbfit{x}_i)}{|\mathbfit{x}-\mathbfit{x}_i|^2}.
    \label{eq:lenseq2}
\end{equation}

We use the surface mass density and shear parameters of QSO 2237+0305 from \citet{2010ApJ...712..658P} for our simulations. These values are given in Table \ref{tab:image_parameters}.

\begin{table}
    \centering
    \caption{Convergence and shear for the macro-images of QSO 2237+0305.}
    \begin{tabular}{|c|c|c|c|c|}
    \hline
    image & A & B & C & D\\
    \hline
    $\kappa$ & 0.40 & 0.38 & 0.73 & 0.62\\
    \hline
    $\gamma$ & 0.40 & 0.39 & 0.72 & 0.62\\
    \hline
    \end{tabular}
    \label{tab:image_parameters}
\end{table}

\begin{figure}
    \centering
    \includegraphics[width=0.5\textwidth]{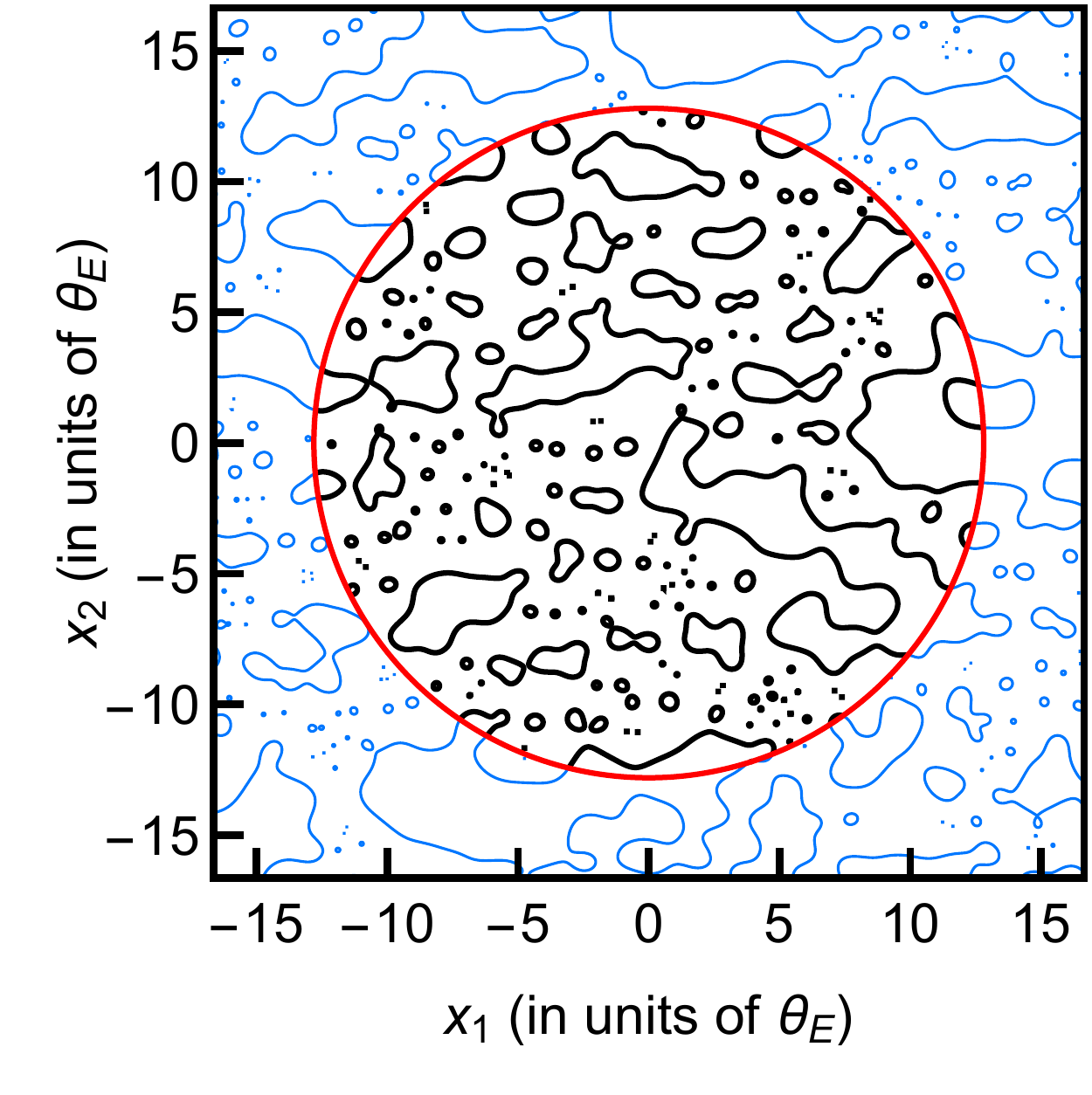}
    \caption{The critical curves for image C ($\kappa=0.73$, $\gamma=0.72$) that were used in our simulations are shown as solid black lines, while the thinner blue lines are those of the surrounding network that were not used. The solid red circle is the region within which the critical curves we considered lie.}
    \label{fig:cc}
\end{figure}

\begin{figure*}
    \centering
    \includegraphics[width=\textwidth]{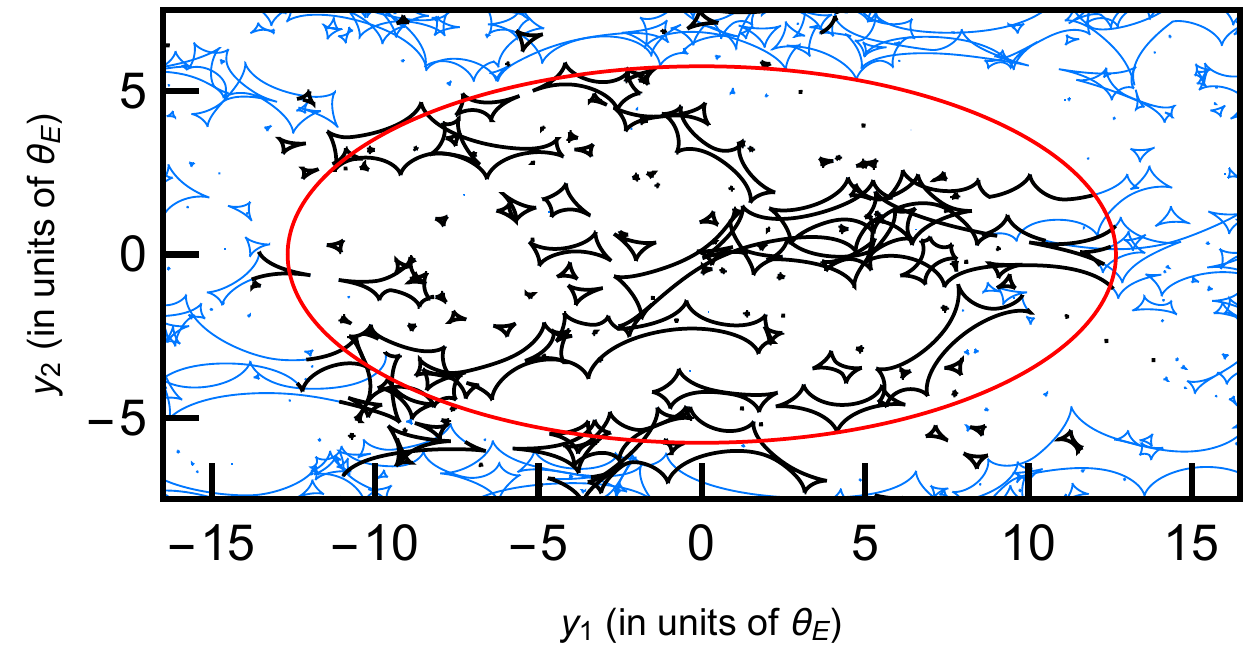}
    \caption{The caustics for image C ($\kappa=0.73$, $\gamma=0.72$) from which we drew our samples are shown as solid black lines, while the thinner blue lines are those of the surrounding network that were not used. The solid red ellipse is the smoothed-out matter mapping of the circle in Fig. \ref{fig:cc}. The caustics considered lie for the most part within this ellipse.}
    \label{fig:caustics}
\end{figure*}

We spread $\approx1000$ stars within a circular region for each case, and use the parametric representation of the critical curves from \citet{1990A&A...236..311W} to precisely locate the critical curves. We consider only those critical curves that lie within a smaller region (to minimize edge effects due to asymmetries in the shear from stars), and their corresponding caustics found by mapping through the lens equation. The critical curves and caustics for the parameters of image C can be seen in Figs. \ref{fig:cc} and \ref{fig:caustics} respectively.

\section{Examination of micro-image magnifications}
\label{sec:micro_image_mags}

We randomly create lines through the source plane to represent different paths that the source might move along. We determine the intersection of these lines with the caustics, and use the intersection points as seeds for our study (using enough lines to generate $\approx50$ seed points). We additionally find the corresponding critical curve locations. The direction of increasing inverse magnification at each point along the critical curves, $\nabla\det\mathbfss{A}$ (where the Jacobian of the lens mapping $\mathbfss{A}=\frac{\partial\mathbfit{y}}{\partial\mathbfit{x}}$ is the inverse magnification matrix, and $\det\mathbfss{A}=0$ defines the locus of critical curves), is both normal to the critical curve and tells us which side of the critical curve the newly created micro-minimum will lie on. Rotating $\nabla\det\mathbfss{A}$ counterclockwise by $\frac{\pi}{2}$ gives a tangent to the critical curve, and mapping this vector to the source plane with $\mathbf{A}$ gives a tangent to the caustic. Further applying a $-\frac{\pi}{2}$ rotation to the resulting vector then gives the direction which is both inside and normal to the caustic. 

With this information, and some further knowledge and manipulation of the lens equation near critical points \citep{1992grle.book.....S}, for each pair of critical curve/caustic seed points we know:

\begin{enumerate}
    \item the source position that induces the creation of a new micro-image pair,
    \item the direction normal to and inside the caustic that we want to follow the source,
    \item which sides of the critical curve the newly created micro-minimum and micro-saddle lie on,
    \item the distance from critical curve to each newly-created image for a source offset normal from the caustic.
\end{enumerate}

We follow a source that moves along a normal direction from one of the caustic seed points, tracking the positions of the newly created micro-images also as they emerges from the corresponding critical curve seed point. The magnifications of the micro-images are calculated at each step, and stored with the source-caustic distance. We note here that we could have tracked the source as it moved along the random lines which we created to determine our seed positions. We are interested in the behavior of the magnifications compared to certain approximations however, and so we choose to follow the source along the normal direction from the caustic.

For the macro-saddles (images C and D), the micro-minima must have a finite `lifetime', and the tracking process is terminated when they annihilate. This is not necessarily true for the macro-minima (images A and B), and in the process of our simulations there were cases where micro-minima appeared abnormally long-lived. We chose to ignore these cases, thus restricting ourselves to always examining micro-minima which eventually annihilated.\footnote{There were a small number of cases where the micro-minima in images A and B survived for lifetimes of order (in the notation of Section \ref{sec:micro_image_stats}) $L=k^2\cdot 10^4-k^2\cdot 10^5$. We chose to toss out such samples.} In all cases, the process for a micro-saddle terminates when the micro-saddle either annihilates or has a magnification less than $10^{-3}$. 

Kayser and Witt provide a simplified form of \citeauthor{1984A&A...130..157C}'s 1984 formula for the caustic strength $K$, giving  

\begin{equation}
    K=\sqrt{\frac{2}{|\mathbfit{T}_{\zeta}|}}
    \label{eq:str_1}
\end{equation} where

\begin{equation}
    \mathbfit{T}_{\zeta}=\mathbfss{A}\cdot \begin{pmatrix}0&-1\\1&0\end{pmatrix}\cdot\nabla_{\mathbfit{x}}\det\mathbfss{A}=\mathbfss{A}\begin{pmatrix}-\frac{\partial\det\mathbfss{A}}{\partial\mathbfit{x}_2}\\\frac{\partial\det\mathbfss{A}}{\partial\mathbfit{x}_1}\end{pmatrix}
    \label{eq:str_2}
\end{equation} is the tangential vector of the caustic \citep{1989A&A...221....1K,1990A&A...236..311W}. 

However, $K$ is the strength for the combined flux of both the created micro-minimum and micro-saddle. We are interested in the behavior of individual images, for which the strength is simply $K/2$ \citep{1992grle.book.....S} that we shall designate here as lowercase $k$.

We calculate the single image caustic strength $k$ at each of our seed critical curve positions, and scale our distances from the caustic by $k^2$. We compare magnification versus scaled distance to the straight fold caustic approximation $\mu=k/\sqrt{d}$ of equation (\ref{eq:std-approx}). 

\begin{figure*}
    \centering
    \includegraphics[width=\textwidth]{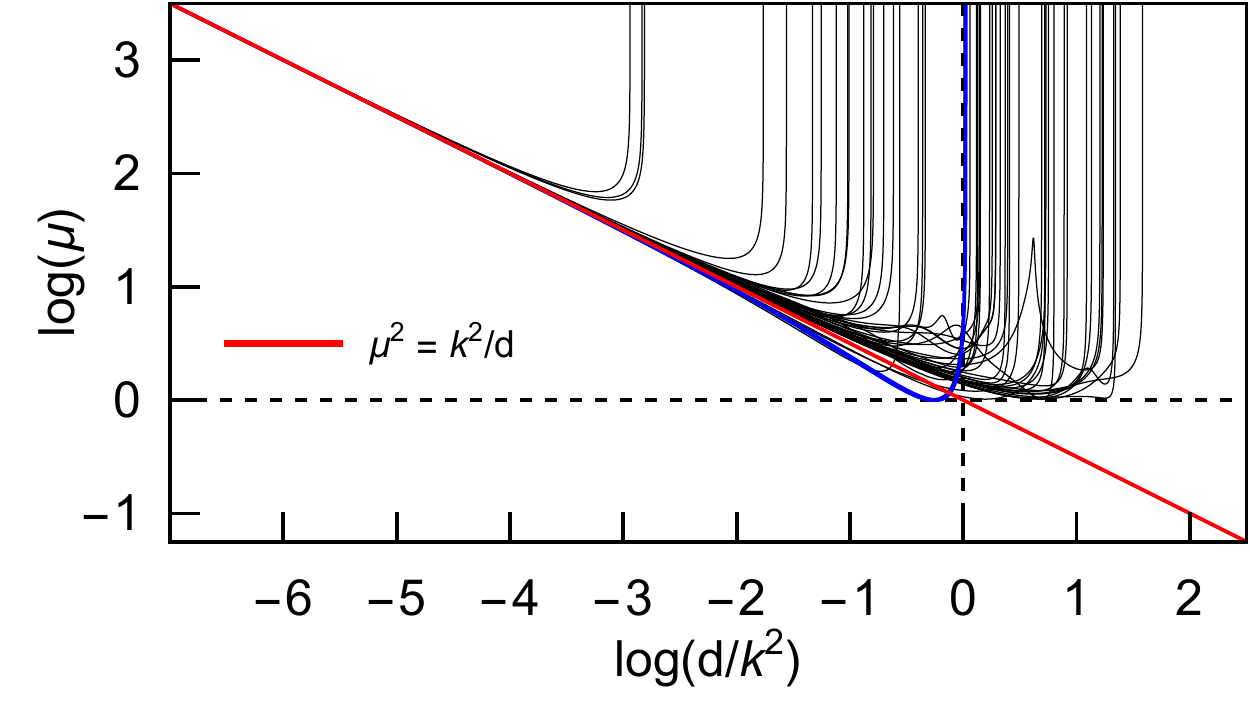}
    \includegraphics[width=\textwidth]{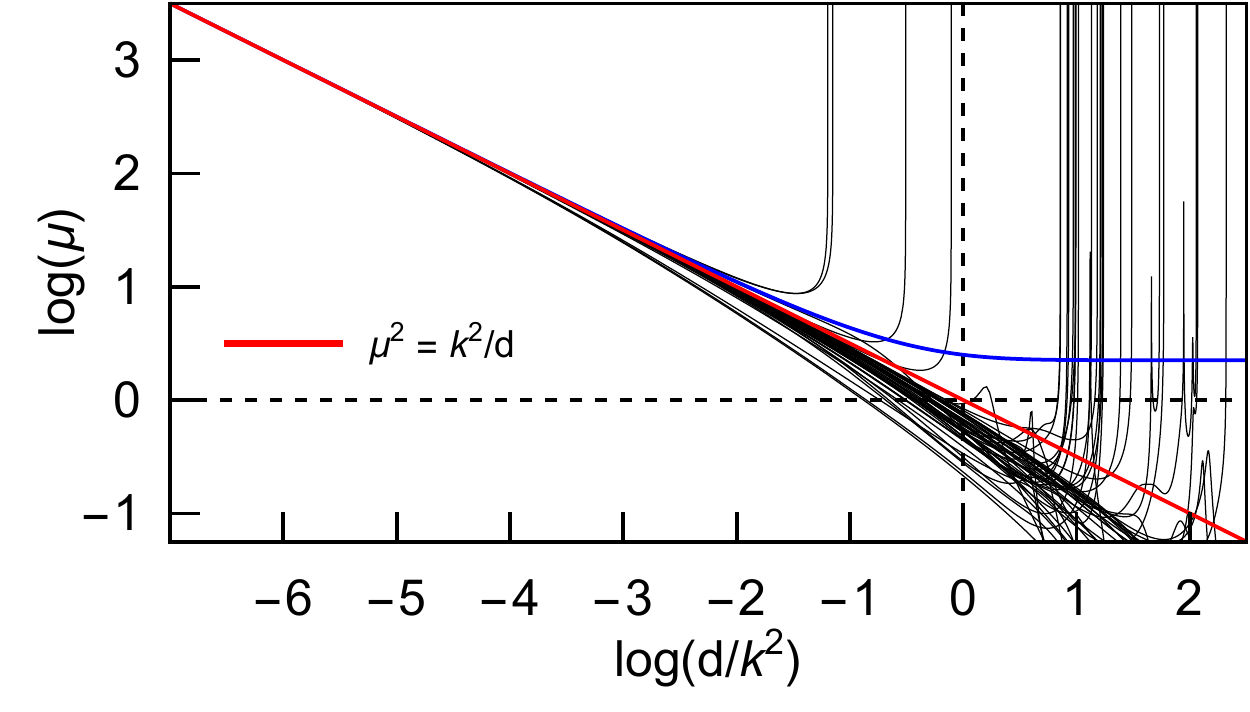}
    \caption{Magnification vs. distance normal from the caustic (scaled by caustic strength). The solid red line shows the standard straight fold approximation of equation (\ref{eq:std-approx}), $\mu=k/\sqrt{d}$. The black lines represent the (top) micro-minima and (bottom) micro-saddles of image C whose magnifications we tracked as they emerged from a critical curve. The blue line is the magnification of the (top) micro-minimum and (bottom) micro-saddle that emerges when a source moves up the symmetry axis of the `deltoid' caustic for a single-star perturbing a macro-saddle of the same magnification as image C. Note that for this single-star perturber, the minimum eventually annihilates at the cusp of the deltoid, and the saddle asymptotically becomes the macro-image.}
    \label{fig:mag_vs_distance}
\end{figure*}

Log-log plots of the results for image C can be seen in Fig. \ref{fig:mag_vs_distance}. The results for images A, B, and D display similar qualitative features. Our results show deviations of the magnifications of micro-images from the approximation $\mu=k/\sqrt{d}$ (shown in the figure as a solid red line) at a distance of $d=k^2$, with noticeable deviation appearing as early as $\log d/k^2=-1$. 

We additionally show (as a blue line) the magnification of the micro-images that appear when a source moves up the symmetry axis of the `deltoid' caustic for a single star perturbing a macro-saddle \citep{1979Natur.282..561C, 1984A&A...132..168C}.\footnote{We chose a macro-saddle with the same macro-magnification as image C. For $\kappa_{\star}=0$, this requires $\gamma>1$.} It is worth pointing out that for our single star perturber, the micro-minimum typically decayed \textit{faster} than in the high stellar density case, and the micro-saddle decayed \textit{slower}. 

\section{Statistics on the `lifetimes' of micro-minima}
\label{sec:micro_image_stats}

Our examination of the magnifications of micro-minima gives us information as well about the typical `lifetime', which we designate by $L$, of a minimum (in units of $k^2$ for its particular point of creation). We additionally examine the lowest magnifications $\mu_{low}$ that the minima reach, and the distance to lowest magnification $d_{low}$ (again rendered dimensionless by $k^2$). 

Again as noted in Section \ref{sec:micro_image_mags}, for images C and D, the micro-minima must have a finite lifetime, while this is not strictly true for images A and B. For images A and B, we only consider those minima which displayed a finite lifetime.

Fig. \ref{fig:mag_vs_distance} shows that the micro-saddles may decay down to very low magnifications, or meet another micro-minimum and annihilate. We do not present statistics on the lifetimes of those micro-saddles which annihilated, but note that for image C half of the micro-saddles decayed to very small magnifications, and half were later annihilated.

\begin{figure}
    \centering
    \includegraphics[width=0.5\textwidth]{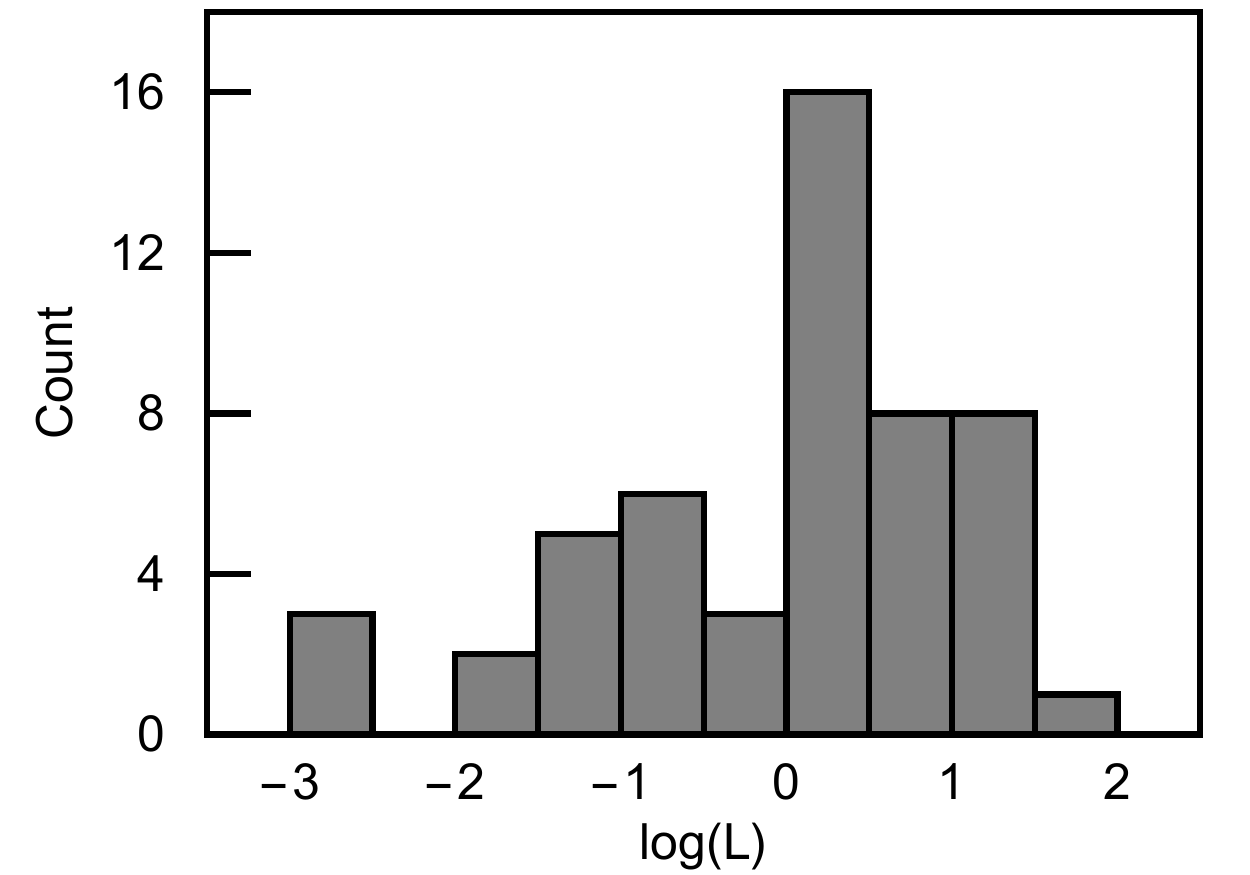}
    \caption{Histogram of logarithms of micro-minima lifetimes $L$ for image C.}
    \label{fig:log_lifetime_histogram}
\end{figure}

\begin{figure}
    \centering
    \includegraphics[width=0.5\textwidth]{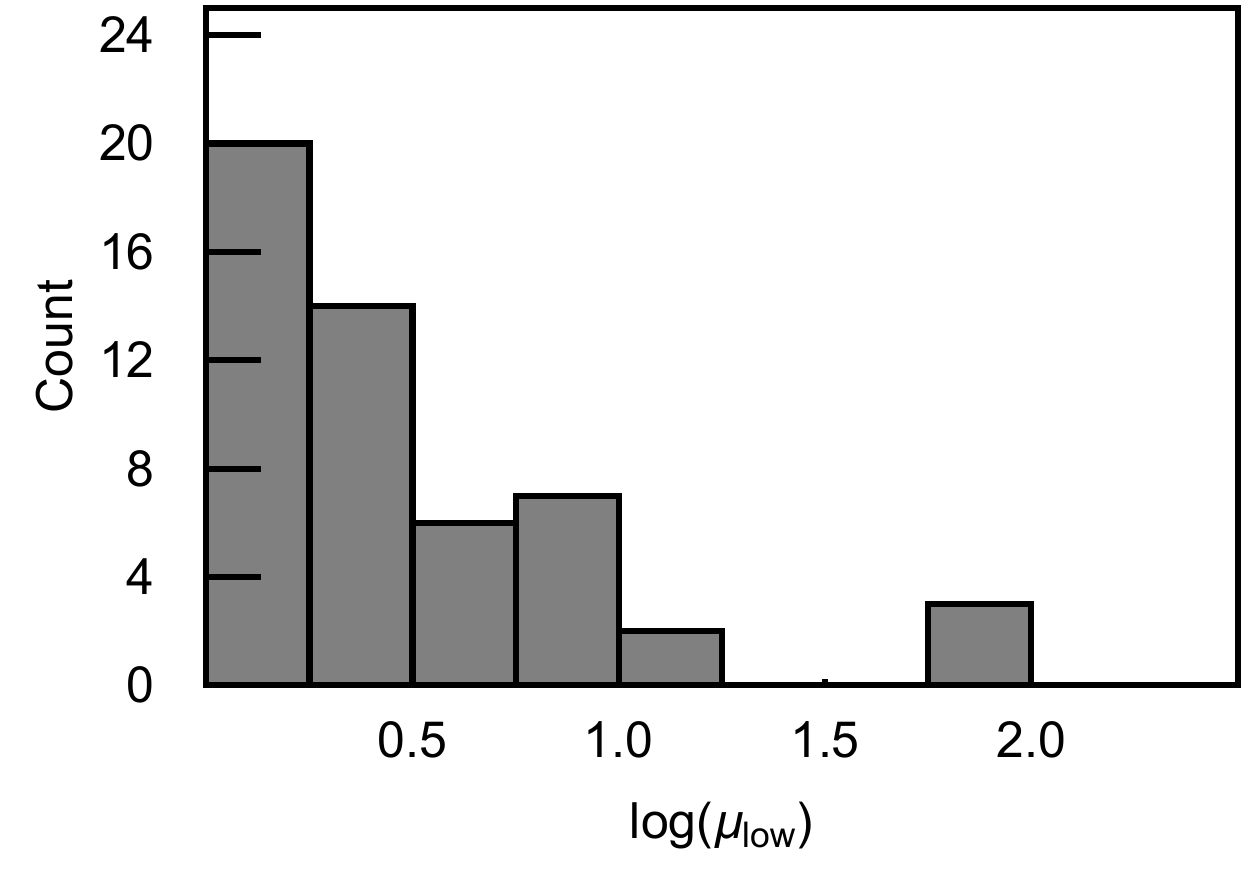}
    \caption{Histogram of logarithms of lowest micro-minima magnifications $\mu_{low}$ for image C.}
    \label{fig:log_lowest_min}
\end{figure}

Fig. \ref{fig:log_lifetime_histogram} gives a histogram of $\log L$ for the micro-minima of image C. We present values for the median and mean of $L$ in Table \ref{tab:min_life_stats}. Fig. \ref{fig:log_lowest_min} shows a histogram of the logarithm of $\mu_{low}$ for image C. We provide values for the median and mean of $\mu_{low}$ and $d_{low}$ in Table \ref{tab:min_mag_stats}.

\begin{table}
    \centering
    \caption{Statistics for the lifetime $L$ (rendered dimensionless by $k^2$) of micro-minima.}
    \begin{tabular}{|c|c|c|c|c|}
    \hline
    image & A & B & C & D\\
    \hline
    Median $L$ & 16.561 & 5.517 & 1.562 & 3.535\\
    \hline
    $\langle L \rangle$ & 26.393 & 13.159 & 5.042 & 8.813\\
    \hline
    \end{tabular}
    \label{tab:min_life_stats}
\end{table}

\begin{table}
    \centering
    \caption{Statistics for the lowest magnifications $\mu_{low}$ of micro-minima.}
    \begin{tabular}{|c|c|c|c|c|}
    \hline
    image&\multicolumn{2}{|c|}{A}&\multicolumn{2}{|c|}{B}\\
    \hline
    $var$ & $\mu_{low}$ & $d_{low}$ & $\mu_{low}$ & $d_{low}$\\
    \hline
    Median $var$ & 1.031 & 6.361 & 1.128 & 2.264\\
    \hline
    $\langle var \rangle$ & 2.173 & 11.522 & 1.644 & 2.970\\
    \hline
    \hline
    image&\multicolumn{2}{|c|}{C}&\multicolumn{2}{|c|}{D}\\
    \hline
    $var$ & $\mu_{low}$ & $d_{low}$ & $\mu_{low}$ & $d_{low}$\\
    \hline
    Median $var$ & 2.130 & 0.896 & 1.269 & 1.734\\
    \hline
    $\langle var \rangle$ & 6.919 & 1.964 & 3.960 & 5.115\\
    \hline
    \end{tabular}
    \label{tab:min_mag_stats}
\end{table}

\section{Distributions of caustic strength}
\label{sec:caustic_strengths}

\begin{figure}
    \centering
    \includegraphics[width=0.4\textwidth]{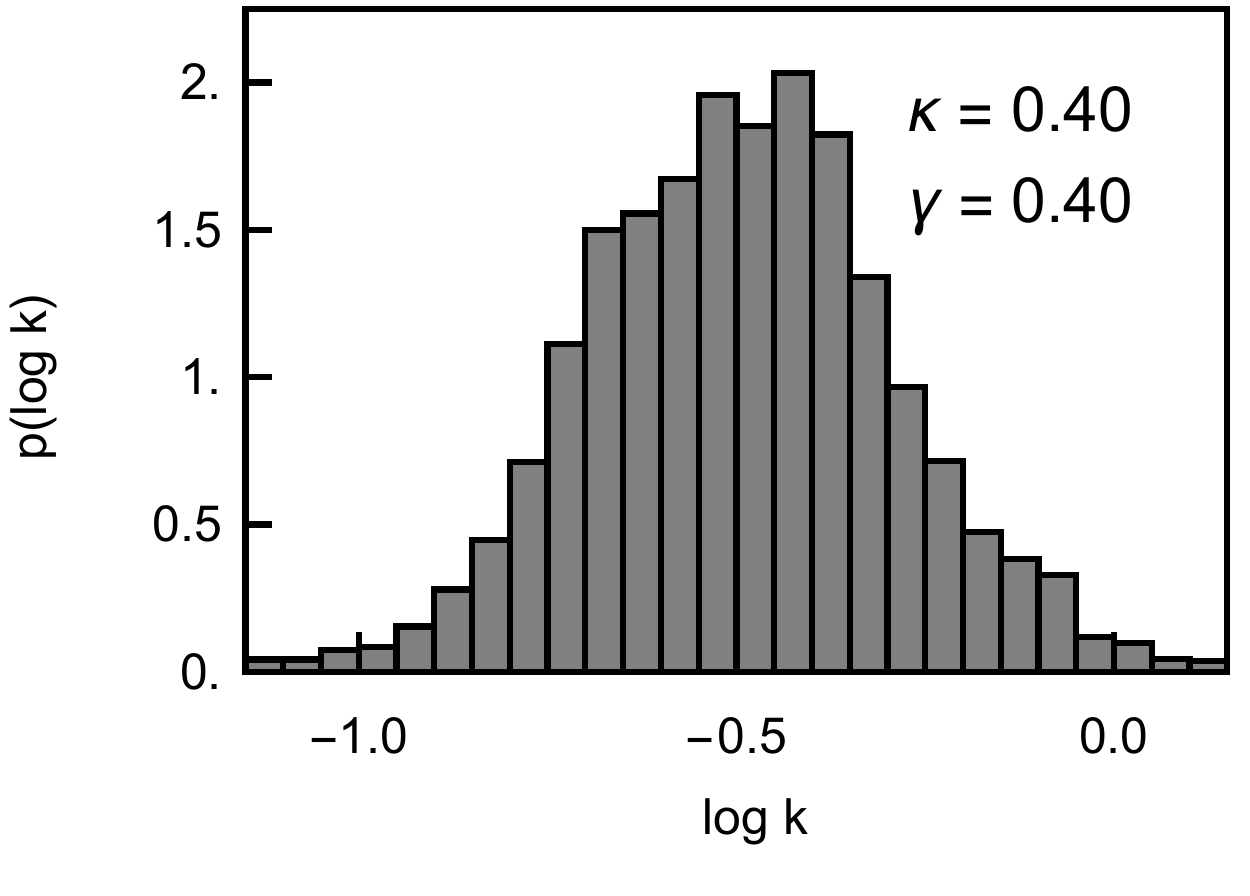}
    \includegraphics[width=0.4\textwidth]{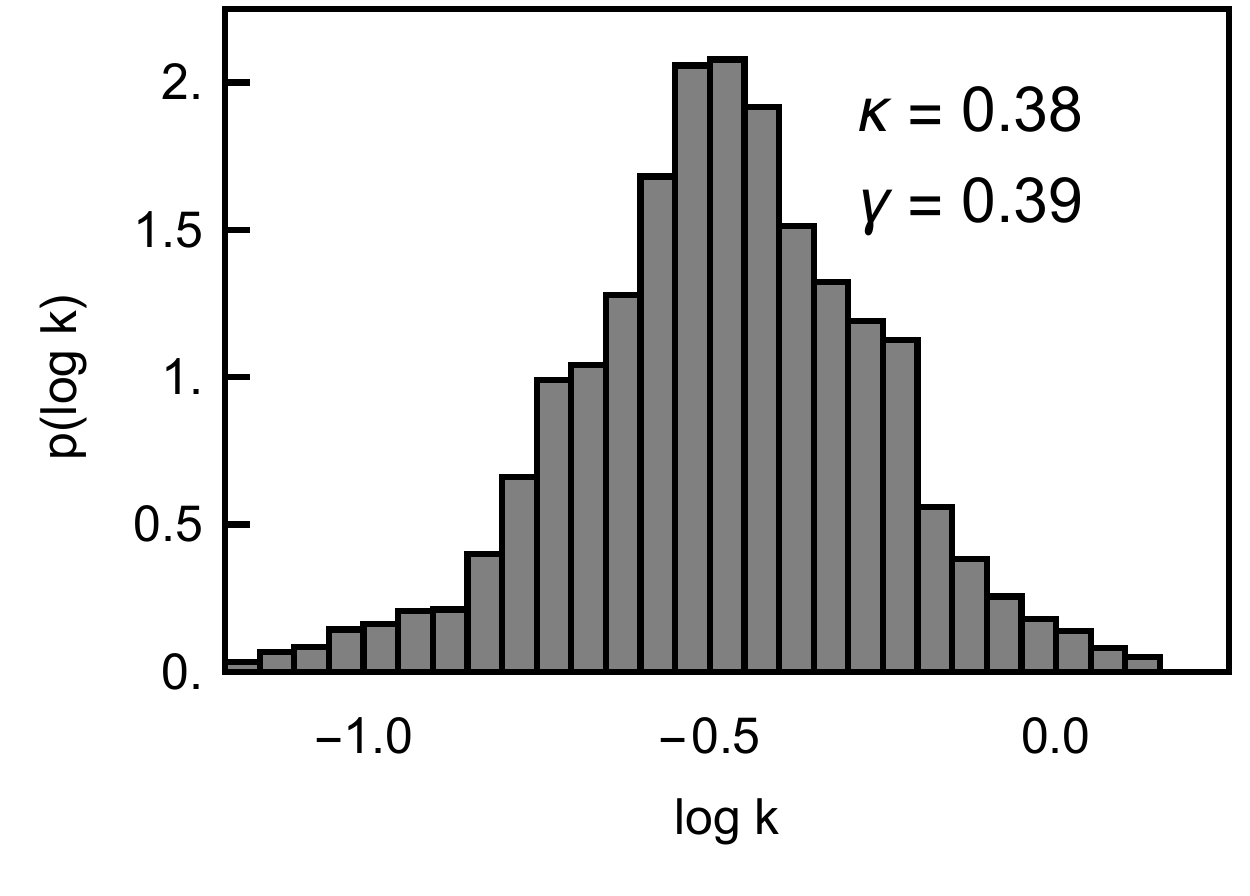}
    \includegraphics[width=0.4\textwidth]{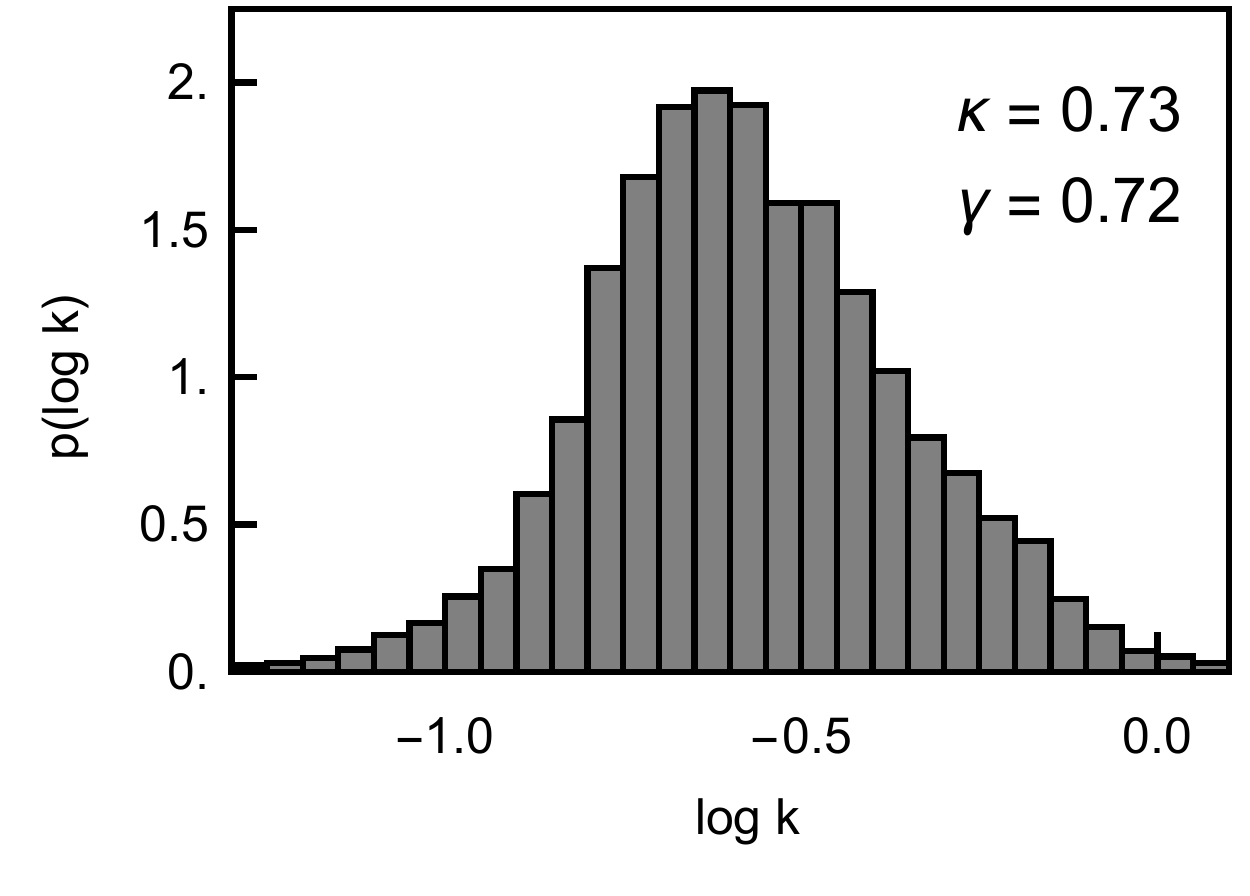}
    \includegraphics[width=0.4\textwidth]{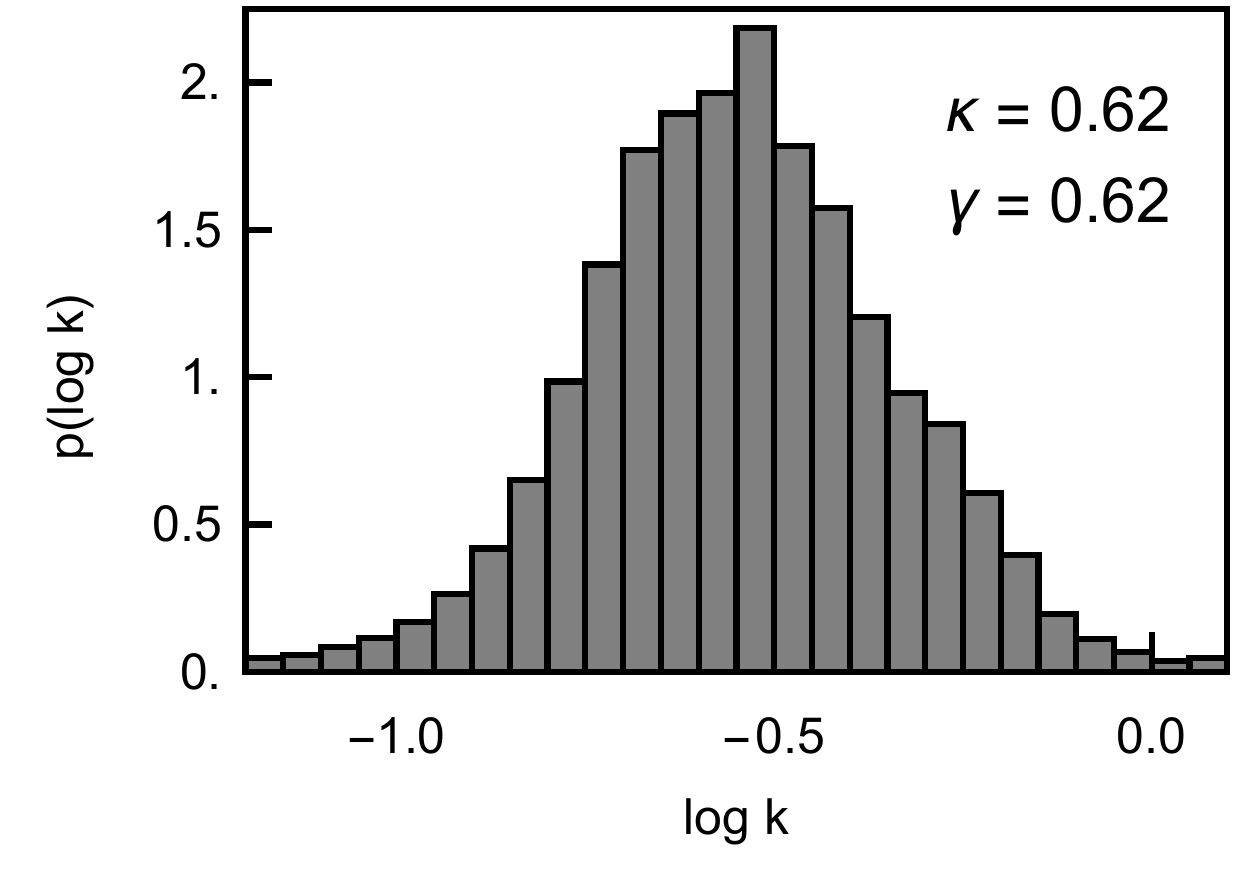}
    \caption{Caustic strength probability density $p(\log k)$ for (from top to bottoom) image A ($\kappa=0.40=\gamma$), image B ($\kappa=0.38$, $\gamma=0.39$), image C ($\kappa=0.73$, $\gamma=0.72$), and image D ($\kappa=0.62=\gamma$). We show $\langle\log k\rangle\pm 3\sigma_{\log k}$.}
    \label{fig:probability_density}
\end{figure}

Witt's parametric representation of the critical curves discretizes the critical curves (and hence the caustics) into sets of points that make polygons which (for appropriately small step sizes of some parameter) appear smooth, as in Figs. \ref{fig:cc} and \ref{fig:caustics}. For every point we found along the caustics, we calculate the single image caustic strength $k$ at the corresponding critical curve location. Under the assumption that $k$ can be considered constant over the small caustic length interval between neighboring points, we can calculate the probability density $p(k)$ for the caustic strength in the source plane. We chose to calculate $p(\log k)$ as well, shown in Fig. \ref{fig:probability_density}. 

\begin{table}
    \centering
    \caption{Statistics of the caustic strength $k$ for QSO 2237+0305.}
    \begin{tabular}{|c|c|c|c|c|}
    \hline
    image & A & B & C & D\\
    \hline
    $\langle k \rangle$ & 0.364 & 0.382 & 0.306 & 0.320\\
    \hline
    $\langle k^2\rangle$ & 0.188 & 2.578 & 0.138 & 0.142\\
    \hline
    $\sigma_k$ & 0.235 & 0.334 & 0.210 & 0.200\\
    \hline
    $\sigma_{k^2}$ & 13.955 & 5.805 & 1.269 & 2.406\\
    \hline
    $\langle \log k\rangle$ & -0.493 & -0.482 & -0.575 & -0.549\\
    \hline
    $\sigma_{\log k}$ & 0.212 & 0.230 & 0.224 & 0.214\\
    \hline
    \end{tabular}
    \label{tab:str_stats}
\end{table}

For the parameters of QSO 2237+0305, we calculate a mean value $\langle k\rangle$, along with $\langle k^2\rangle$ and $\sigma_k=\sqrt{\langle k^2\rangle-\langle k\rangle^2}$. We also calculated $\langle k^4\rangle$, but only for the purpose of finding $\sigma_{k^2}=\sqrt{\langle k^4\rangle-\langle k^2\rangle^2}$. Our results are presented in Table \ref{tab:str_stats}, along with the mean and standard deviation for $\log k$. 

\begin{figure*}
    \centering
    \includegraphics[width=\textwidth]{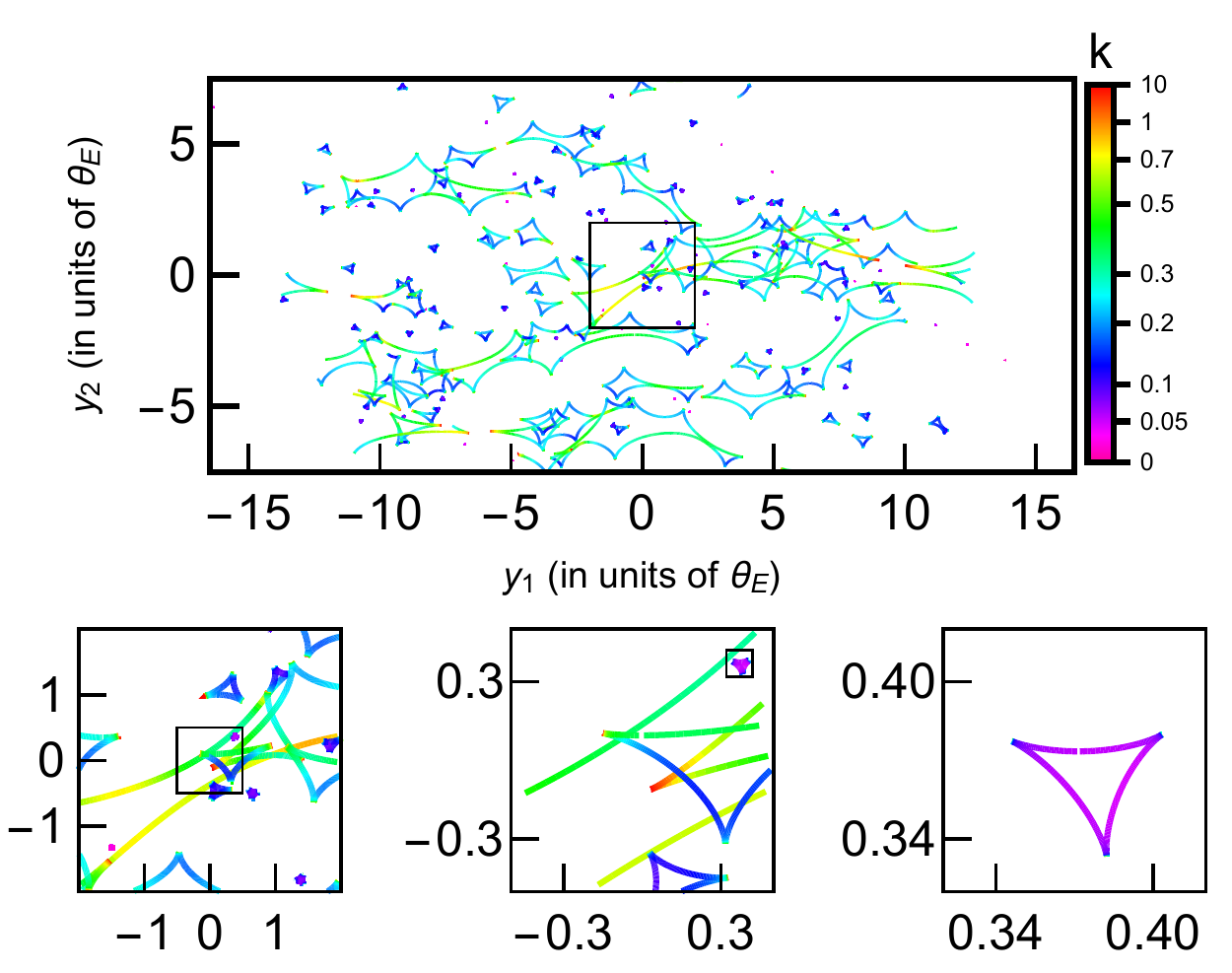}
    \caption{The portion of the caustic network of image C that was used to calculate $p(k)$ (i.e. the solid black caustics in Fig. \ref{fig:caustics}) has been color coded according to the caustic strength at each point. Subsequent zooms of regions are shown in the bottom row.}
    \label{fig:caustic_strength_color_plot}
\end{figure*}

While Fig. \ref{fig:probability_density} provides one look at the distribution of caustic strengths for image C, we show as well in Fig. \ref{fig:caustic_strength_color_plot} the caustic network of image C where each point has been color-coded by the value of the caustic strength $k$.

\citet{1990A&A...236..311W} has calculated distributions of the caustic strength for the case of zero external shear and a range of surface mass density values corresponding to macro-minima and maxima. \citet{1998JKAS...31...27L} have calculated distributions of the caustic strength for a binary lens with external shear. \citet{1993A&A...268..501W} provide values of $\langle K\rangle$ for (slightly different parameters of) QSO 2237+0305 as well, though they do not show the underlying distributions. The authors are unaware of anywhere in the literature where distributions of the caustic strength for values of surface mass density and shear corresponding to saddlepoints are given. 

\section{Higher order magnification approximations}
\label{sec:higher_order_approx}

\citet{2005ApJ...635...35K} and \citet{2011MNRAS.417..541A} provide higher order approximations for the magnifications of images near critical curves. Through a Taylor expansion of the lens equation in the vicinity of a critical curve that produces a fold caustic, \citet{2005ApJ...635...35K} derived \begin{equation}
    \label{eq:keeton_et_al_approx}
    \mu_{\pm}^{-1}=\pm a\cdot\sqrt{d}+b\cdot d
\end{equation}where the plus and minus symbols denote the signed magnifications of minima and saddles respectively, $d$ is normal to the caustic, \begin{equation}
    a=\sqrt{2\tau_{22}^2\tau_{111}}=\frac{1}{k},
\end{equation}and \begin{equation}
    b=\frac{2}{\tau_{111}}\Big(\frac{1}{3}\tau_{22}\tau_{1111}-\tau_{112}^2+\tau_{111}\tau_{122}\Big).
\end{equation}

The variable $\tau=\frac{1}{2}(\mathbfit{x}-\mathbfit{y})^2-\psi(\mathbfit{x})$ is the gravitational time delay \citep{1986ApJ...310..568B}. Subscripts denote derivatives with respect to the first or second coordinate of the image plane, evaluated at the origin (around which we take the Taylor expansion of the lens equation to occur). We have written $a$ and $b$ such that at the origin, our caustic normal points along the abscissa axis of some coordinate system.\footnote{This is a rotation of the coordinate system by $-\pi/2$ from that which appears in \citet{2005ApJ...635...35K} and \citet{2011MNRAS.417..541A}. This was chosen to be consistent with the normal and tangent vectors from \citet{1990A&A...236..311W}.} In order to evaluate $a$ and $b$, we can first calculate all the derivatives up to order 4 at each critical curve seed in the global coordinate system of eq. (\ref{eq:lenseq2}) that aligns with the external shear.\footnote{Many simplifications arise due to commutativity of derivatives and the fact that $\tau_{11}+\tau_{22}=2$ for our point mass lens model.} We then rotate our coordinate basis to a local system at each seed caustic point where the normal points along the abscissa axis, and determine the necessary derivatives present in $a$ and $b$.

Results from \citet{2011MNRAS.417..541A} are in agreement with \citet{2005ApJ...635...35K}, and additionally contain a next order term. We do not present the resulting lengthy equations here, but note that the next order terms contain not only the distance normal to the caustic, but also the distance tangential.\footnote{However, as we only examine a source moving normal to the caustic, this tangential distance does not come into play for our analysis. This is also why do not dwell on the `parabolic fold caustic' approximation of \citet{1999MNRAS.302...68F}, as along the normal direction it is equal to that of the straight fold caustic.} We perform a similar process as that described above for $a$ and $b$ in order to find the coefficients present in their approximation.

\begin{figure*}
    \centering
    \includegraphics[width=\textwidth]{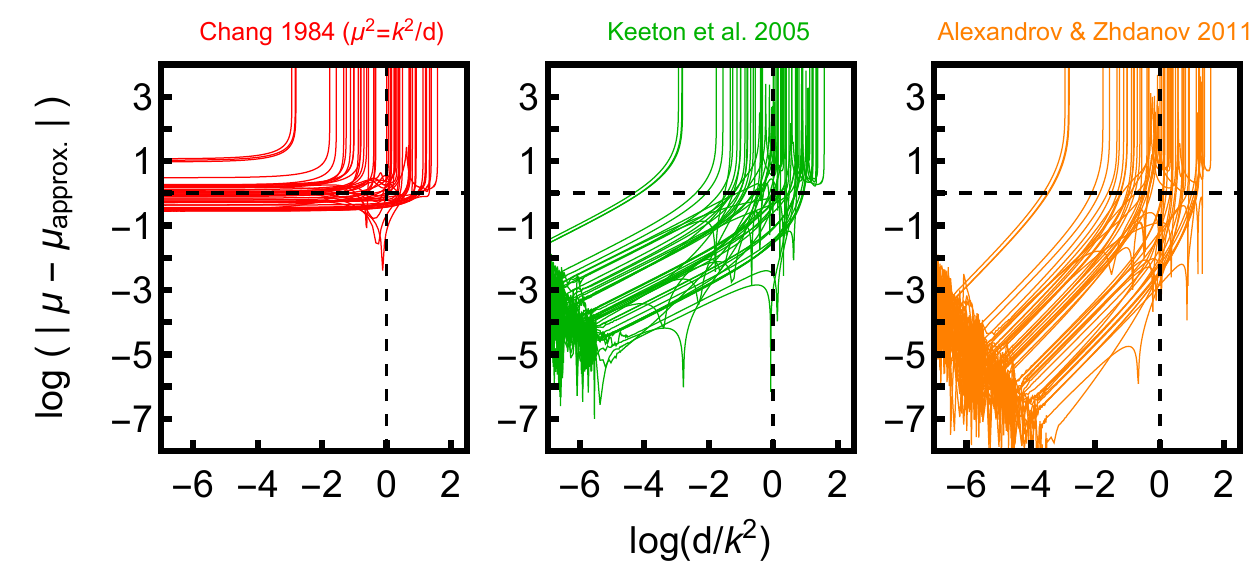}
    \caption{Differences between actual magnification of micro-minima in image C and predictions for three different approximations.}
    \label{fig:error_plot}
\end{figure*}

We can compare the actual magnifications $\mu$ of the micro-images in our simulations with the predictions $\mu_{approx.}$ of these two higher order approximations. We present the results for micro-minima in Fig. \ref{fig:error_plot}, and note that errors for the micro-saddles display similar behavior. 

The middle and right plots of Fig. \ref{fig:error_plot} show a lot of scatter in the error for small values of $d/k^2$. This is likely due to the numerical precision of our simulations very close to the critical curves -- however, the error in magnification is small (of order $10^{-4}-10^{-5}$) when the actual magnification itself is very large (of order $10^3$)! There are large downward spikes visible where the error goes to $0$. As in the top of Fig. \ref{fig:mag_vs_distance} where the blue curve of the micro-minimum intersects the straight red line around $\log(d/k^2)=-0.1$, the approximation formally gives the exact magnification, though this is only due to the difference changing sign from positive to negative. Positions where the error goes to infinity are locations where either the micro-minimum annihilates, or the approximation gives infinite magnification. The approximation may become infinite before or after the minimum annihilates, depending on the values and signs of the coefficients in the approximation. If after, there is only one infinite spike in the error (when the minimum annihilates). If before, multiple spikes (ones when the approximation goes to infinity, and the final one when the minimum annihilates).

In general, Fig. \ref{fig:error_plot} makes it clear that including the next higher order terms for the magnification reduces the error significantly for small values of $d/k^2$, as is expected. However, none of the approximations are consistently well suited to the regime where $d/k^2\approx 1$ for the micro-minima.

\section{Statistics for a higher order approximation}
\label{sec:higher_order_approx_stats}

Much like how one can calculate the distribution of $k$ along the caustics, we can do so as well for $a$ and $b$ in the approximation \begin{equation}
    \mu_\pm^{-1}=\pm a\cdot\sqrt{d}+b\cdot d.
\end{equation} We provide the results for these calculations in Table \ref{tab:higher_order_coeff_stats}, including $a^2$ as well. We can then consider how an `average' micro-image might behave with these parameters.

\begin{table}
    \centering
    \caption{Statistics for coefficients of eq. \ref{eq:keeton_et_al_approx} for QSO 2237+0305.}
    \begin{tabular}{|c|c|c|c|c|}
    \hline
    image & A & B & C & D\\
    \hline
    $\langle a \rangle$ & 3.516 & 3.511 & 4.301 & 4.029\\
    \hline
    $\langle a^2\rangle$ & 16.665 & 17.855 & 26.036 & 23.343\\
    \hline
    $\sigma_a$ & 2.075 & 2.351 & 2.745 & 2.666\\
    \hline
    $\sigma_{a^2}$ & 55.969 & 123.591 & 245.028 & 243.325\\
    \hline
    $\langle b\rangle$ & -8.134 & -6.963 & -13.031 & -12.041\\
    \hline
    $\sigma_b$ & 1383.166 & 1027.542 & 521.262 & 3677.847\\
    \hline
    \end{tabular}
    \label{tab:higher_order_coeff_stats}
\end{table}

We include $\langle a^2\rangle$ because of possible choices one might make to non-dimensionalize distances: one can take $d\cdot\langle a\rangle^2$, or $d\cdot\langle a^2\rangle$.\footnote{One could also take $d\cdot\langle b\rangle$, though this seems less useful to the authors.} Numerically, there might be slight differences based upon this choice. We provide results for the various combinations resulting from each choice in the following discussions. 

The fact that we found $\langle a\rangle$ and $\langle b\rangle$ to be of opposite sign suggests that a micro-minimum with \begin{equation}
    \label{eq:avg_keeton_et_al_approx}
    \mu_+^{-1}=\langle a\rangle\sqrt{d}+\langle b\rangle d
\end{equation} has an effective mean lifetime $\langle d\rangle$ of \begin{equation}
    \langle d\rangle=\Big(-\frac{\langle a\rangle}{\langle b\rangle}\Big)^2,
\end{equation} as $\mu^{-1}=0$ at this distance.\footnote{We note that there is not \textit{always} such an effective lifetime, as the actual value of $a$ and $b$ along the critical curve may be of the same sign.} This distance is in units of $\theta_E$, but can be rendered into a dimensionless lifetime \begin{equation}
    \label{eq:approx_lifetime}
    \langle L\rangle=a^2\langle d\rangle 
\end{equation} (where one might choose $\langle a\rangle^2$ or $\langle a^2\rangle$ as the actual multiplier for $\langle d\rangle$) to compare with the results of Section \ref{sec:micro_image_stats}, Table \ref{tab:min_life_stats}. There are 3 unique combinations for the variables in Table \ref{tab:higher_order_coeff_stats} that provide such a dimensionless lifetime, and their results are presented in Table \ref{tab:approx_mean_lifetimes}.

\begin{table}
    \centering
    \caption{Mean micro-minimum lifetime $\langle L\rangle$ (rendered dimensionless by $a$) for QSO 2237+0305.}
    \begin{tabular}{|c|c|c|c|c|}
    \hline
    image & A & B & C & D\\
    \hline
    $\frac{\langle a\rangle^2}{\langle b\rangle^2}\langle a\rangle^2$ & 2.310 & 3.134 & 2.015 & 1.817\\
    \hline
    $\frac{\langle a\rangle^2}{\langle b\rangle^2}\langle a^2\rangle$ & 3.114 & 4.540 & 2.836 & 2.614 \\
    \hline
    $\frac{\langle a^2\rangle}{\langle b\rangle^2}\langle a^2\rangle$ & 4.198 & 6.575 & 3.992 & 3.758\\
    \hline
    \end{tabular}
    \label{tab:approx_mean_lifetimes}
\end{table}

Similarly, one can find that the minimum magnification of eq. (\ref{eq:keeton_et_al_approx}) (if $a$ and $b$ are of opposite sign) is \begin{equation}
    \label{eq:approx_mu_low}
    \mu_{low} = -\frac{4b}{a^2} 
\end{equation} and occurs at a dimensionless \begin{equation}
    \label{eq:approx_d_low}
    d_{low} = \frac{a^4}{4b^2}.
\end{equation} Values for $\langle\mu_{low}\rangle$ are given in Table \ref{tab:approx_mean_lowest_min} for the choices of $\langle a^2\rangle$ and $\langle a\rangle^2$. Values for $\langle d_{low}\rangle$ are easily found from Table \ref{tab:approx_mean_lifetimes} since $\langle d_{low}\rangle=\frac{1}{4}\langle L\rangle$.

\begin{table}
    \centering
    \caption{Mean micro-minima lowest magnification $\langle \mu_{low}\rangle$ for QSO 2237+0305.}
    \begin{tabular}{|c|c|c|c|c|}
    \hline
    image & A & B & C & D\\
    \hline
    $-\frac{\langle a\rangle^2}{4\langle b\rangle}$ & 2.632 & 2.257 & 2.818 & 2.967\\
    \hline
    $-\frac{\langle a^2\rangle}{4\langle b\rangle}$ & 1.952 & 1.560 & 2.263 & 2.063\\
    \hline
    \end{tabular}
    \label{tab:approx_mean_lowest_min}
\end{table}

The approximations for $\langle L\rangle$ and $\langle d_{low}\rangle$ from eqs. (\ref{eq:approx_lifetime}) and (\ref{eq:approx_d_low}) respectively are in general lower than the results of Section \ref{sec:micro_image_stats}. This gives values of $\langle\mu_{low}\rangle$ slightly higher than in Section \ref{sec:micro_image_stats}. Altogether, this is a reminder then that once a minimum begins to move away from its point of creation, the presence of other caustics begins to play a more important role in its behavior. This makes its behavior highly unpredictable based solely off the few parameters one might ascertain when the minimum comes into being. The qualitative result from our simulations seems to be that micro-minima tend on average to drop in magnification slower than what one expects from the leading order approximation $\mu=k/\sqrt{d}$, but slightly faster than one might anticipate from any predictions based on higher order approximations. 

\begin{figure*}
    \centering
    \includegraphics[width=\textwidth]{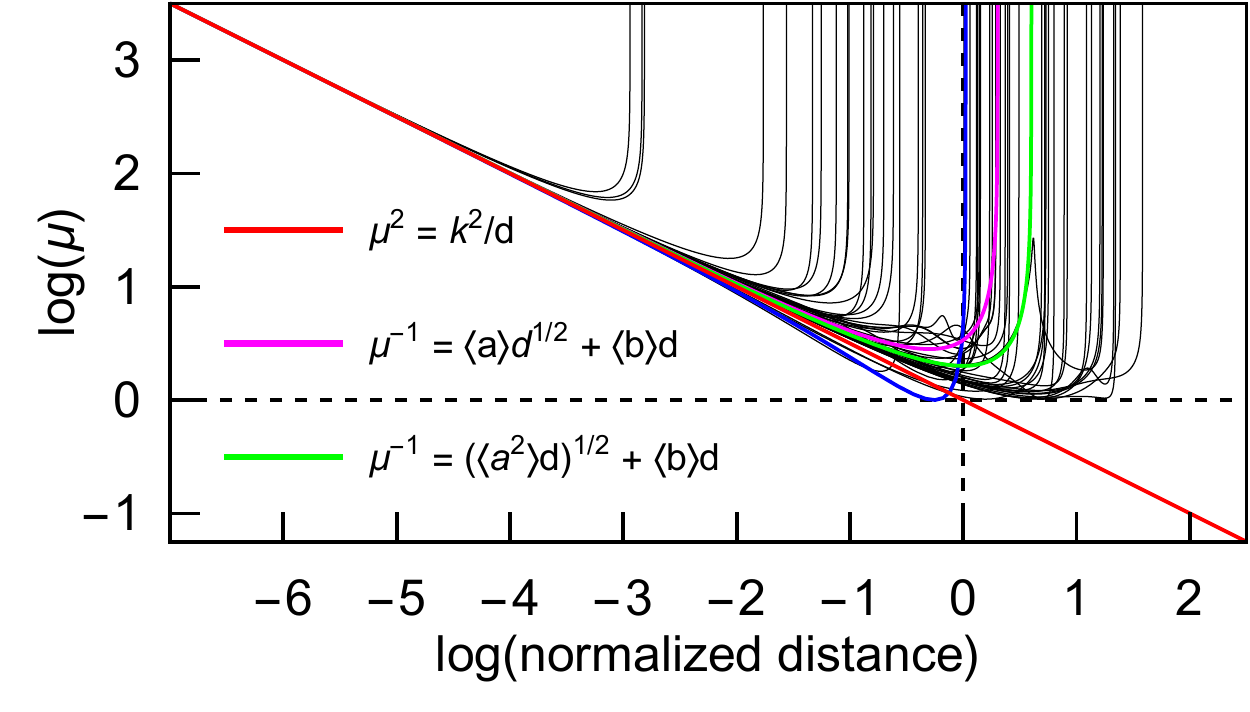}
    \includegraphics[width=\textwidth]{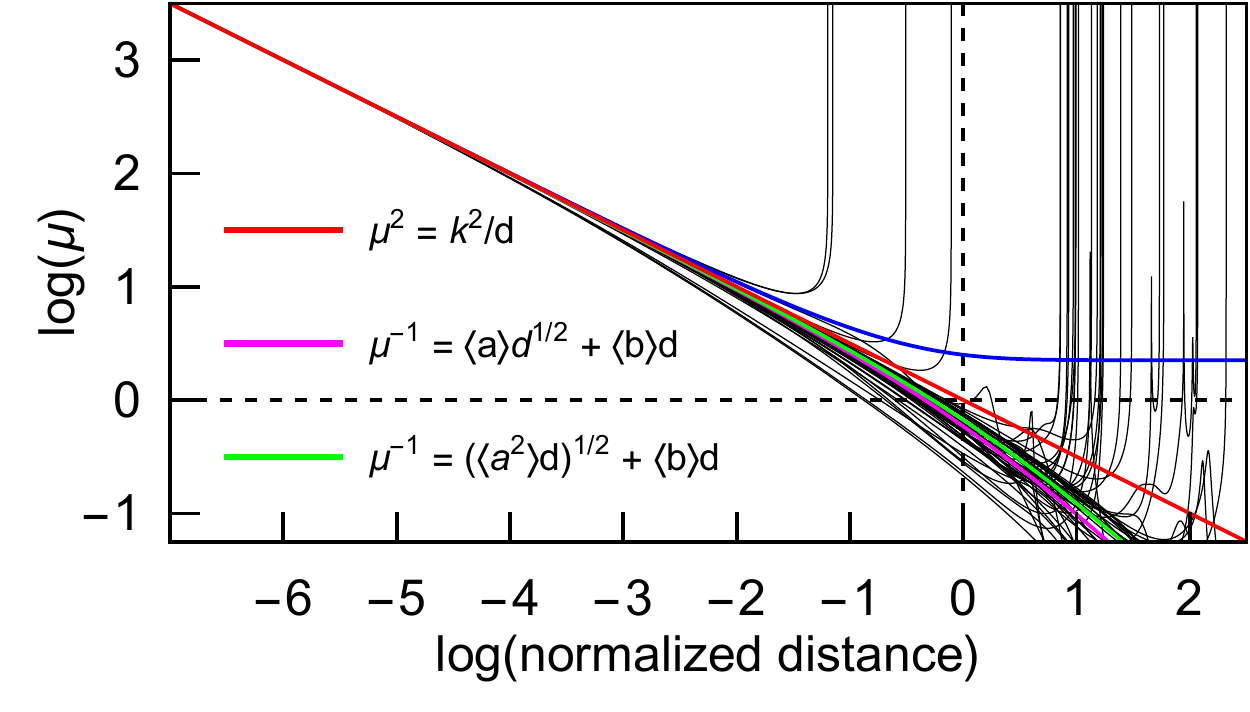}
    \caption{We show again the magnifications of the micro-minima and micro-saddles of image C, with the addition of the next order approximation of eq. (\ref{eq:keeton_et_al_approx}) for coefficients with the average values indicated from Table \ref{tab:higher_order_coeff_stats}.}
    \label{fig:mag_vs_distance_with_approx}
\end{figure*}

Finally, we also show in Fig. \ref{fig:mag_vs_distance_with_approx} the two higher approximations $\mu_\pm^{-1}=\pm\langle a\rangle\sqrt{d}+\langle b\rangle d$ and  $\mu_\pm^{-1}=\pm\sqrt{\langle a^2\rangle d}+\langle b\rangle d$ for image C, along with the actual micro-image magnifications. Much as we scale the distances for the micro-images by using the local value of $k$, we scale the distances for these two approximations by the appropriate local value of $k=1/\langle a\rangle$ or $k^2=1/\langle a^2\rangle$ respectively. 

\section{Shape profiles of caustic crossing events with a higher order approximation}
\label{sec:alt_approx}

Many analyses of caustic crossing events designed to examine the size of the light emitting region of an AGN rely on the convolution of a source luminosity profile with the magnification approximation near fold caustics of equation (\ref{eq:std-approx}). \citet{1987A&A...171...49S} present details of the shapes such caustic events take in the lightcurves for two example luminosity profiles. We briefly examine here how higher order approximations to the magnification might affect these shapes.

We take as our source a uniform circular disc with luminosity profile $L(x,y)=H(R^2-(x-x_s)^2-(y-y_s)^2)$, where $R$ is the radius of the source, $(x_s,y_s)$ is the center of the source, $H$ is the Heaviside step function, and $x$ and $y$ are axes in the source plane. 

\begin{figure}
    \centering
    \includegraphics[width=0.5\textwidth]{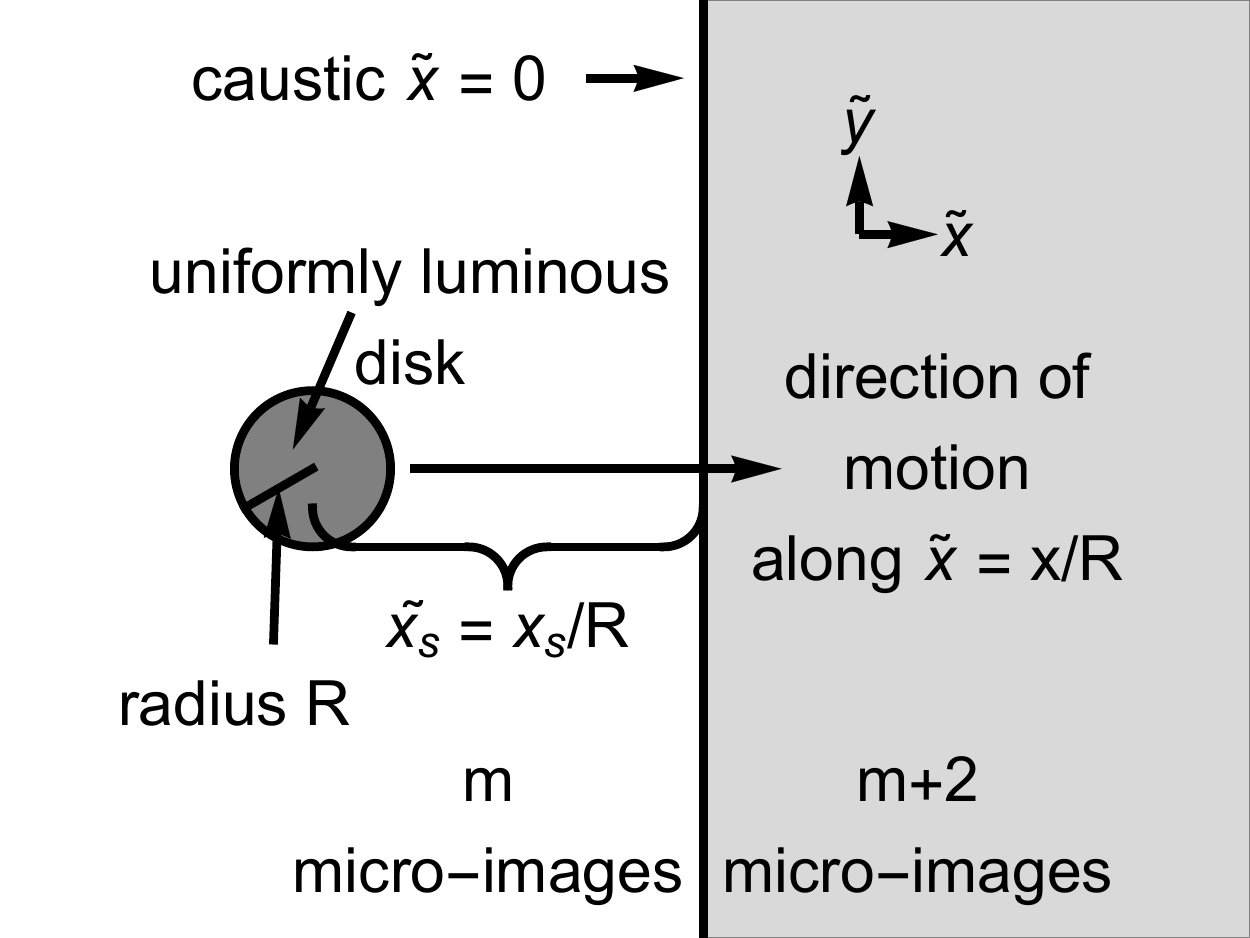} 
    \caption{Visualization for the setup of a uniform disc crossing a fold caustic in the normal direction from outside (lower magnification, $m$ micro-images) to inside (higher magnification, $m+2$ micro-images).}
    \label{fig:uniform_disk_setup}
\end{figure}

We take our caustic to be $x=0$. We use the approximation for the magnification \begin{equation}
    \mu(x,y)=\mu_{minimum}+\mu_{saddle}=\frac{1}{a\sqrt{x}+bx}+\frac{1}{a\sqrt{x}-bx}.
\end{equation} We assume the source crosses the caustic in the normal direction, and so we can further take $y_s=0$ for our source without loss of generality.\footnote{We have also ignored the additional flux from other existing micro-images. Such a term is generally assumed to be slowly-varying over the length scales of interest, contributing an additive constant that can be pulled out of the integrals.} Fig. \ref{fig:uniform_disk_setup} provides a visualization of this description.  The magnification of our source is then \begin{equation}
\begin{aligned}
    &\frac{\iint L(x,y)\mu(x,y)dxdy}{\iint L(x,y)dxdy}=
    \\&\frac{1}{\pi R^2}\iint H(R^2-(x-x_s)^2-y^2)\cdot\Big(\frac{1}{a\sqrt{x}+bx}+\frac{1}{a\sqrt{x}-bx}\Big)dxdy
\end{aligned}
\end{equation}where the integrals are taken over the entire source plane.

Integrating over $y$, we arrive at \begin{equation}
    \frac{2}{\pi R^2}\int_{Max(0,x_s-R)}^{Max(0,x_s+R)} \sqrt{R^2-(x-x_s)^2}\Big(\frac{1}{a\sqrt{x}+bx}+\frac{1}{a\sqrt{x}-bx}\Big)dx.
\end{equation}

With only the leading order approximation $\mu=2k/\sqrt{d}$ of eq. (\ref{eq:std-approx}), the caustic strength is simply an overall scale factor that can be pulled out of the integral, thus determining the maximum magnification and having nothing to do with the shape. That is no longer the case now however -- some assumption about the values of $R$, $a$, and $b$ must be made to proceed.

For the sake of our example, we take $a=\sqrt{\langle a^2\rangle}$ and $b=\langle b\rangle$ for the parameters of image C. We then choose three values of the source size, $R\in\{0.03\theta_E, 0.01\theta_E, 0.003\theta_E\}$. We can then plot magnification $\mu$ vs. source-caustic distance $x_s$, where we choose to scale the distance by the value of the source radius. 

\begin{figure}
    \centering
    \includegraphics[width=0.5\textwidth]{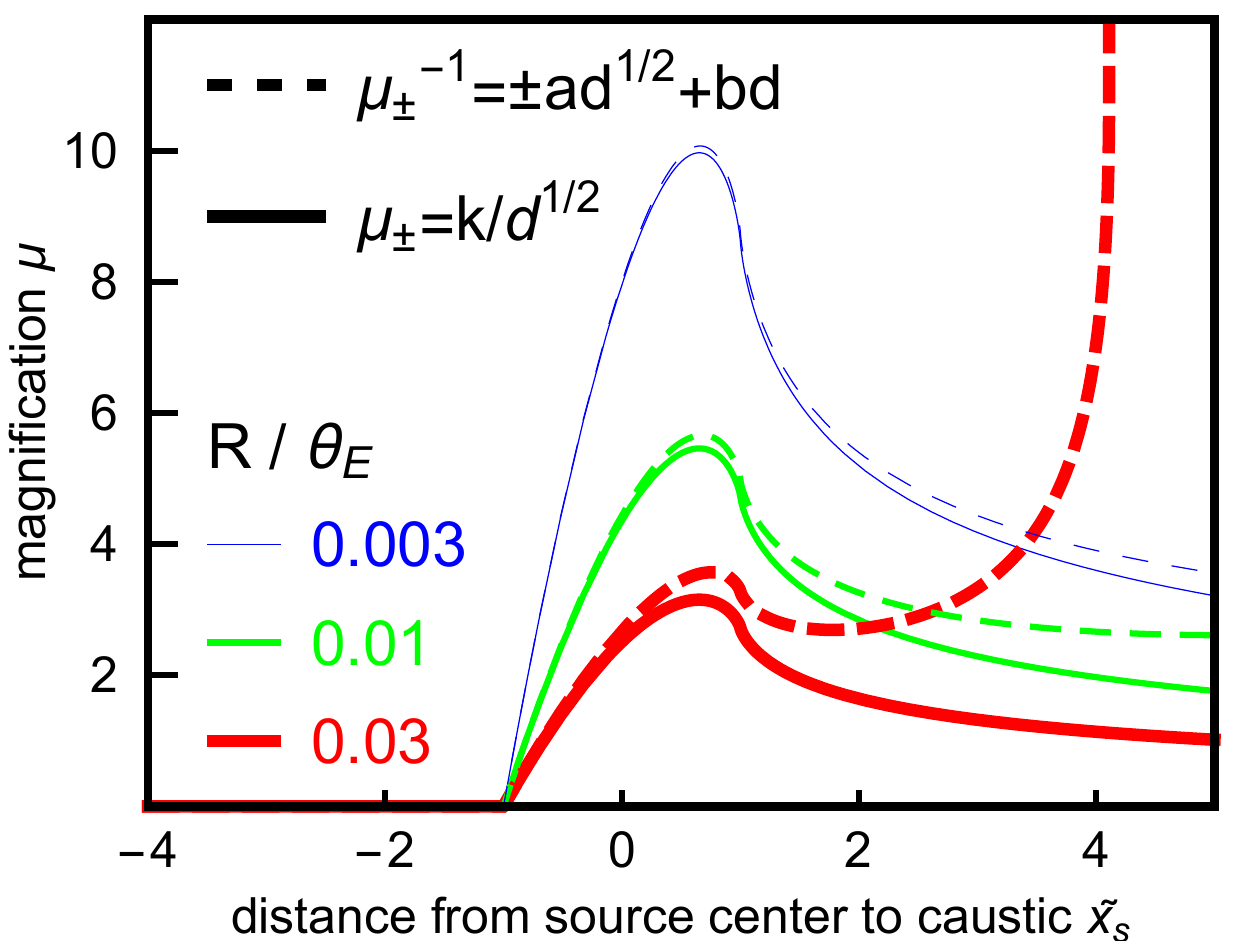} 
    \caption{Magnification vs. source center distance from caustic for various values of $R$. Solid lines indicate the standard approximation of eq. (\ref{eq:std-approx}), while dashed lines include the next order term, eq. (\ref{eq:keeton_et_al_approx}). We have used mean values from image C, $a=\sqrt{\langle a^2\rangle}$ and $b=\langle b\rangle$. Compare with Fig. \ref{fig:mag_vs_distance_with_approx} where the solid red line signifies the standard approximation of equation (\ref{eq:std-approx}) and the solid green line includes the higher order correction eq. (\ref{eq:keeton_et_al_approx}).}
    \label{fig:uniform_disk_magnification}
\end{figure}

The resulting magnification curves as a function of $\Tilde{x_s}=x_s/R$ for our various values of $R$ can be seen in Fig. \ref{fig:uniform_disk_magnification}. The higher value of $R=0.03\theta_E$ means that once inside the caustic, the source covers a significant portion of the region within the `effective' lifetime of the minimum, $d=\frac{\langle a^2\rangle}{\langle b\rangle^2}=0.153\theta_E$ for image C. It quickly annihilates, and the prohibitively large increase in magnification towards the end is a reminder of the failings of the approximation at larger distances. For the smaller values of $R$, there would be similar behavior at larger values of $\Tilde{x_s}$ outside our plotted range. However, the source has time to settle down in magnification before reaching these locations. We note that for our selected values of $a$ and $b$, the source never goes below a magnification of $1$, as is required since a micro-minimum is present. For the standard approximation however, given enough distance the approximation would provide a lower magnification than is allowed.

In general, the higher order approximation provides a higher peak magnification than the old, though the difference becomes less noticeable as $R$ decreases. Additionally, with the approximation from eq. (\ref{eq:std-approx}) the magnification profile always peaks at the same value of $\Tilde{x_s}=2/3$ \citep{1987A&A...171...49S}. This is no longer the case for the higher order, as the peak occurs at a value of $\Tilde{x_s}>2/3$ and appears to approach $\Tilde{x_s}=2/3$ as $R$ decreases.

\section{Conclusions}
\label{sec:conclusions}

We have examined the magnifications of micro-images near fold caustics, and found that their magnifications differ from the inverse square root approximation typically used as the standard. We find that significant differences occur at distances $d$ equal to the square of the caustic strength $k$, with noticeable deviations appearing as early as $\log d/k^2=-1$. 

We have presented as well some statistics on the behavior of the lifetimes and lowest magnifications of the micro-minima in our simulations. Additionally, we provide probability distributions of the caustic strengths for the macro-images of QSO 2237+0205 (Huchra's lens). 

We have compared the actual magnifications of the micro-images in our simulations to the higher order approximations of \citet{2005ApJ...635...35K} and \citet{2011MNRAS.417..541A}. We find that including the next higher order terms can greatly reduce the error for small values of $d$, but provides little help in the regime where $\mu\approx 1$. We include statistics on values of parameters appearing in one such higher order approximation. 

Additionally, we examine the effect that a higher order approximation has on the `shape profiles' for a source crossing a fold caustic. In general, the peak of the curve occurs at a smaller source-caustic distance, and with a higher peak magnification.

\section*{Acknowledgements}
This work was supported by the MIT Undergraduate Research Opportunities Program and the Deutsch-Amerikanische Fulbright-Kommission. We thank the anonymous referee for their comments, which led to significant improvements in this paper.




\bibliographystyle{mnras}
\bibliography{references.bib} 


\bsp	
\label{lastpage}
\end{document}